\documentclass{emulateapj}

\def\sles{\lower2pt\hbox{$\buildrel {\scriptstyle <}
   \over {\scriptstyle\sim}$}}
\def\sgreat{\lower2pt\hbox{$\buildrel {\scriptstyle >}
   \over {\scriptstyle\sim}$}}

\def \bc{\begin{center}}
\def \ec{\end{center}}

\shorttitle{Curvature Effect of a Non-powerlaw Spectrum } \shortauthors{Zhang et
al. }
\begin{document}

\title{Curvature Effect of a Non-Power-Law Spectrum and Spectral Evolution of
GRB X-Ray Tails}

\author{Bin-Bin Zhang\altaffilmark{1}, Bing Zhang \altaffilmark{1},
  En-Wei Liang \altaffilmark{2}, Xiang-Yu Wang \altaffilmark{3} 
} \altaffiltext{1}{Department of Physics and Astronomy, University of Nevada,
Las Vegas,
NV 89154, USA; zbb,bzhang@physics.unlv.edu}
\altaffiltext{2}{Department of Physics, Guangxi University, Nanning 530004,
China}
\altaffiltext{3}{Department of Astronomy, Nanjing University, Nanjing 210093,
China}
\begin{abstract}
The apparent spectral evolution observed in the steep decay
phase of many GRB early afterglows raises a great concern of the
high-latitude ``curvature effect'' interpretation of this phase. 
However, previous curvature effect models only invoked a simple power 
law spectrum upon the cessation of the prompt internal emission. We 
investigate a model that invokes the ``curvature effect'' of a more
general non-power-law spectrum and test this model with
the Swift/XRT data of some GRBs. We show that one can
reproduce both the observed lightcurve and the apparent spectral 
evolution of several GRBs using a model invoking a power-law spectrum
with an exponential cut off. GRB 050814 is presented as an example.
\end{abstract}


\section{INTRODUCTION\label{sec:intro}}
Most of the early X-Ray afterglows detected by Swift (Gehrels
et al. 2004)
show a steep decay phase around 100$\sim$1000
seconds after the burst trigger (Tagliaferri et al. 2005). 
The main characteristics of this steep decay phase include the
following. (1) It
connects smoothly to the prompt $\gamma$-ray light curve extrapolated 
to the X-ray band, suggesting that it is the ``tail'' of the prompt 
emission (Barthelmy et al. 2005, O'Brien et 2006,
Liang et al 2006).  (2) The decay slope is typically $3 \sim 5$ when
choosing the GRB trigger time as the zero time point $t_0$
(Tagliaferri et al. 2005; Nousek et al. 2006; Zhang et al 2006). 
(3) The time-averaged spectral index of the steep decay phase is
much different from that of the later shallow decay phase, indicating
that it is a distinct new component that is unrelated to the
conventional afterglow components (Zhang et al 2006; Liang et
al. 2007). (4) Strong spectral evolution exists in about one third
of the bursts that have a steep decay phase (Zhang et al. 2007, 
hereafter ZLZ07; Butler \& Kocevski 2007; 
Starling et al. 2008). All these features suggest that
the steep decay phase holds the key to understand the connection
between the prompt emission (internal) phase and the traditional 
afterglow (external) phase. Any proposed model (see M\'esz\'aros
2006; Zhang 2007 for reviews) should be able to 
explain these features.

The so called ``curvature effect'', which accounts for the delayed 
photon emission from high latitudes with respect to the line of 
sight upon the abrupt cessation of emission in the prompt emission
region (Fenimore et al. 1996; Kumar \& Panaitescu 2000; 
Dermer 2004; Dyks et al. 2005; Qin 2008a), 
has been suggested to play an important role in shaping the sharp 
flux decline in GRB tails (Zhang et al. 2006; Liang et al. 
2006; Wu et al. 2006; Yamazaki et al. 2006). In the simplest model, 
it is assumed that 
the instantaneous spectrum at
the end of the prompt emission is a simple power law 
with a spectral index $\beta$. The predicted temporal decay index
of the emission is (with the convention $F_{\nu} \propto t^{-\alpha} 
\nu^{-\beta}$)
\begin{equation}
 \alpha=2+\beta~,
\label{curvature-1}
\end{equation}
if the time origin to define the $\log-\log$ light curve, $t_0$, 
is taken as the beginning of the last emission episode
before the cessation of emission. Adopting a time-averaged
$\beta$ in the tails, Liang et al. (2006) 
found that Eq.(\ref{curvature-1})
is generally valid. 
The strong spectral evolution identified 
in a group of GRB tails (ZLZ07)
apparently violates Eq.(\ref{curvature-1}), which is valid only
for a constant $\beta$. ZLZ07 then investigated a curvature effect
model by assuming a structured jet with varying $\beta$ at different
latitudes and that the line of sight is near the jet 
axis\footnote{Notice
 that this structured jet model is different from the traditional
 one that invokes an angle-dependent energy/Lorentz factor, but
 not the spectral index (Zhang \& M\'esz\'aros 2002; Rossi et al. 2002).}.
One would then expect that Eq.(\ref{curvature-1}) is roughly
satisfied, with both $\alpha$ and $\beta$ being time-dependent.
ZLZ07 found that this model does not fit the data well.

These facts do not rule out the curvature effect interpretation of
GRB tails, however. This is because the instantaneous spectrum upon
the cessation of prompt emission may not be a simple power law.
If the spectrum has a curvature, as the 
emission from progressively higher latitudes reach the observer,
the XRT band is sampling different segments of the intrinsic
curved spectrum (Fig.1). This would introduce an apparent spectral
evolution in the decaying tail. 
The main goal of this paper is to test this more general
curvature effect model using the available Swift XRT data.

\section{Curvature Effect of a Non-powerlaw Spectrum }

We consider a general non-power-law spectrum in the form of 
\begin{equation}
 F_\nu(\nu)=F_{\nu,c}G(\nu)~,
\end{equation}
where $G(\nu)$ is the function form of the spectrum
with a characteristic frequency $\nu_c$ so that $G(\nu_c)$=1, 
and $F_{\nu,c}=F_\nu (\nu_c)$ is the normalization of the 
spectrum at $\nu=\nu_c$. 

The curvature effect states that given a same spectrum
at different latitudes with respect to the line of sight,
one has 
$F_{\nu,c}\propto {\cal D}^2$ and $\nu_c \propto {\cal D}$,
where ${\cal D}$ is the Doppler factor. If the high-latitude
angle $\theta \gg \Gamma$, the Dopper factor ${\cal D}
\propto t^{-1}$, so that $F_{\nu,c} \propto t^{-2}$,
$\nu_c \propto t^{-1}$ (Kumar \& Panaitescu 2000).
Considering the $t_0$ effect (Zhang et al. 2006; Liang
et al. 2006), this can be written as
\begin{equation}\label{fe}
F_{\nu,c}(t) = F_{\nu,c,p}  
\left(\frac{t-t_0}{t_p-t_0}\right)^{-2}
\end{equation}
and 
\begin{equation}\label{ep}
 \nu_c(t) = \nu_{c,p} 
\left(\frac{t-t_0}{t_p-t_0}\right)^{-1}
\end{equation}
for $t \gg t_p$, 
where $t_0$ refers to the time origin of the last pulse in the 
prompt emission and $t_p$ is the epoch when the curvature-effect 
decay starts (or the ``peak'' time of the lightcurve), 
$F_{\nu,c,p}=F_{\nu,c}(t_p)$ and $\nu_{c,p}=\nu_c(t_p)$.
Notice that in the case of $G(\nu)=(\nu/\nu_c)^{-\beta}$
(a pure power law spectrum), one derives $F_\nu \propto
(t-t_0)^{-\beta-2}$. This is the relation Eq.(\ref{curvature-1}).

We consider several physically motivated non-powerlaw spectra with 
a characteristic frequency $\nu_c$, including the cut-off power law 
spectrum and the
Band-function (Band et al. 1993). To explore the compatibility with 
the data, we also investigate different forms of the cutoffs with 
varying sharpness. In all cases, the $F_{\nu_p}(t)$ and
$\nu_p(t)$ follow Eqs.(\ref{fe}) and
(\ref{ep}). When $\nu_c(t)$ drops across an observational narrow 
energy band, e.g. the Swift/XRT band, it introduces an apparent
spectral softening with time, which, if fitted by a power law,
shows an increase of photon index with time.
In the meantime, the flux within the observing band
drops down rapidly, leading to an apparent steep decay phase 
in the lightcurve (Fig.1). 

\begin{table}
  \caption[]{Best-fitting parameters and their 1-sigma errors for 
the cutoff power curvature effect model for GRB050814.  }
\begin{tiny}
  \begin{center}\begin{tabular}{lllllll}
  \hline\noalign{\smallskip}
$N_{0,p}$ & $E_{c,p}({\rm keV})$ & $\Gamma$ & $t_0({\rm s})$ & nH$_{host}$&
k & $\chi^2$/dof\\
  \hline\noalign{\smallskip}
$0.67(0.12)$ &$10.2(1.3)$ &$1.56(0.25)$ &$103.5(3.4)$ &0.002(0.04)&1 (fixed)&$10.7/9$\\
  \hline\noalign{\smallskip}

  \end{tabular}\end{center}
\end{tiny}
\end{table}

\begin{figure}\label{cartoon}
 \includegraphics[angle=0,scale=0.7]{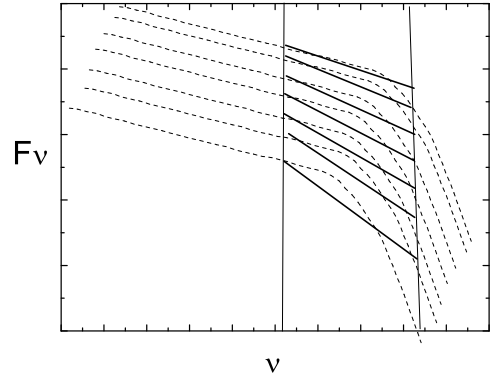}

\caption{A schematic picture showing that shifting a set of
non-power-law spectra in time can equivalently give an apparent spectral 
evolution in a fixed band. The dashed lines represent a set of 
exponential-like spectra, whose $F_{\nu_p}(t)$ and $\nu_p(t)$ 
drop down with time according to Eqs. (\ref{fe}) and (\ref{ep}). 
The two vertical solid lines bracket the observed energy band. 
The thick solid lines denote the effective power law fits
to the time-dependent spectra at each time step.} 
%\\ \end{tiny}
\end{figure}

\begin{figure}\label{050814fitting}
 \includegraphics[angle=0,scale=.45]{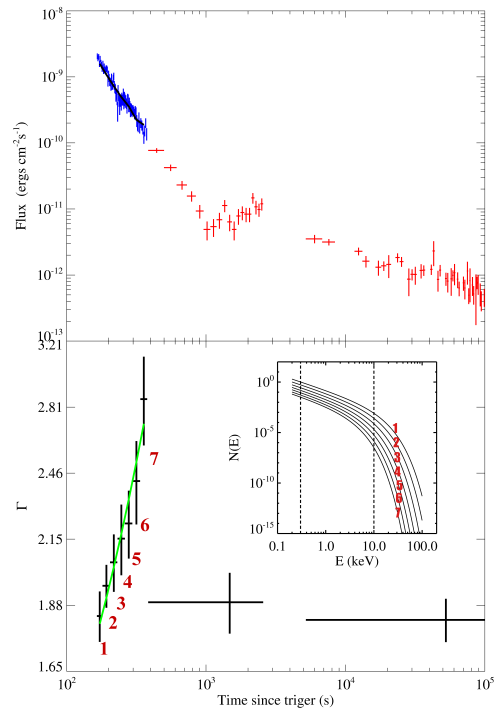}

\caption{The lightcurve (upper panel) and spectral evolution (lower
panel) of the X-ray tail of GRB 050814 with the best-fit theoretical
model (black curve in upper panel and green curve in lower panel).
The blue and red data points are the window
timing and photon counting data, respectively. The inset shows 
time-dependent theoretical spectra with the XRT band (0.3-10 keV)
bracketed by two vertical lines.  The integers denote the 
time segments for the time-resolved spectral analysis.}

\end{figure}

\begin{figure}
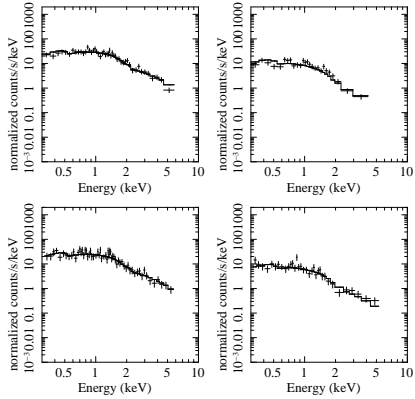

\begin{center}
  \includegraphics[angle=270,scale=0.13]{f3a.ps}
  \includegraphics[angle=270,scale=0.13]{f3c.ps} \\
  \includegraphics[angle=270,scale=0.13]{f3d.ps} 
  \includegraphics[angle=270,scale=0.13]{f3f.ps}
\caption{\textit{Upper panel}: Examples of simulated time-dependent spectra 
of GRB050814 with the best-fit parameters. The time intervals are 
1,6 repectively as denoted in Fig. 2. In each panel, the data histogram
displays the simulated spectrum, and the solid line displays the
best-fit ($\chi^2/dof = 39.0/61,  25.2/25$) power
law model ($wabs*zwasb*powerlaw$ in XSPEC) that is used to derive the
time-dependent photon index $\Gamma$. \textit{Lower pabel}: The
corresponding observed spectrum in the three time intervals and
their power law fits ($\chi^2/dof = 47.1/46, 22.0/19$).}
\end{center}

\end{figure}

\begin{figure}
\begin{center}
 \includegraphics[angle=270,scale=0.25]{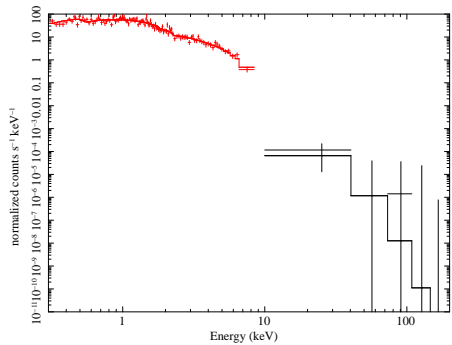}
\end{center}

\caption{The simulated cut-off power law spectrum at $t_p=144$ s 
based on the best fit model confronted by the BAT data in the time 
interval $(141.5-146.5)$ s. The reduced $\chi^2$ of the fitting 
is 1.2 with dof$=197$.}
\end{figure}

\section{Data Reduction and Simulation Method\label{sec:sim}}

We consider a time-dependent cutoff power law photon spectrum taking 
the form of  
\begin{equation}\label{eqmain}
N(E,t)=N_{0}(t) \left(\frac{E}{1~{\rm keV}}\right)^{-\Gamma} 
\exp\left[-\left(\frac{E}{E_c(t)}\right)^k \right]
\end{equation}
where $\Gamma=\beta+1$ is the power law photon index, and
$k$ is a parameter to define the sharpness of the high energy 
cutoff in the spectrum, $E_c(t)$ is the time-dependent 
characteristic photon energy, and $N_0(t)$ is a 
time-dependent photon flux (in units of
${\rm photons\cdot keV^{-1}cm^{-2}s^{-1}}$) at 1 keV (Arnaud 1996). 
The choice of this function was encouraged by the fact that
the spectral evolution of some GRB tails can be fitted by such
an empirical model (Campana et al. 2006; ZLZ07; Yonetoku et al. 2008).
According to Eqs.(\ref{fe}) and (\ref{ep}), and 
noticing the conversion between the photon flux and the emission 
flux density, i.e. $F_\nu \propto EN(E)$,  we get
\begin{equation}\label{Ec}
E_c(t) = E_{c,p} \left(\frac{t-t_0}{t_p-t_0}\right)^{-1}
\end{equation}
and $N(E_c,t) = N_{c,p} \left[(t-t_0)/(t_p-t_0)\right]^{-1}$,
where $N_{c,p}=N(E_c,t_p)$, and $E_{c,p}=E_c(t_p)$. This gives
\begin{equation}\label{Nc_02}
N_{0}(t) = N_{0,p} \left(\frac{t-t_0}{t_p-t_0}\right)^{-(1+\Gamma)}~.
\end{equation}
Notice that $t_p$ is the beginning of the steep decay, which is
a parameter that can be directly constrained by the data. For
a complete lightcurve, we read $t_p$ off from
the lightcurve. In the case of an observational gap, usually 
$t_p$ can be reasonably fixed to the end of the prompt emission. 
We therefore do not include this parameter into the fits, and
derive the other five parameters, namely, $N_{0,p}$, $E_{c,p}$, 
$\Gamma$, $t_0$, and $k$ from the data. At any time $t$, the
model spectrum can be determined once these parameters are 
given. One can then confront the model with the real GRB data. 

The procedure includes the following steps.
(1) For a given burst, we extract its Swift/XRT light curve and $n$
slices of time-dependent spectra using the standard HEASoft/Swift
Package. The details of the data reduction method were described 
in ZLZ07. (2) Given a trial set of parameters in the theoretical 
spectra\footnote{Notice that $k$ is fixed 
to a certain value for a particular model, and is varied when 
different models are explored.} \{$N_{0,p}$, $E_{c,p}$, $\Gamma$, 
$t_0$\}, using Eqs.(\ref{eqmain}-\ref{Ec}) we model $n$
time-dependent {\em theoretical spectra} that correspond to 
the time bins that are used to derive the time-dependent
observed spectra.
(3) Based on the theoretical spectra of each time slice, we 
simulate the corresponding {\em model spectra} by taking account 
of the observational effects, including the Swift/XRT response 
matrix, the absorption column densities ($N_H$) of both the Milky 
Way (extracted from the observations from step 1) and the host galaxy 
of the burst (a free parameter),
the redshift (if applicable), and a Poisson noise background. 
Notice that $n_{H,host}$ is another parameter introduced in
the model spectra (besides the other parameters introduced
in the theoretical spectra). All these faked spectra can
be obtained using HEASoft (Version 6.4) and Xspec (Version 12.4) 
(4) We fit the faked model spectra with a simple power law model, 
i.e. $wabs*wabs*powerlaw$ (or $wabs*zwabs*powerlaw$ if the redsift 
is available) in XSPEC and get the simulated fluxes and spectral
indices of the $n$ slices. Here the column densities of both the 
Milk Way and the host galaxy are fixed to the obsered values as 
in Step 1. (5) We compare the simulated fluxes and spectral indices with the 
observed ones and access the goodness of the fits using $\chi^2$
statistics. (6) We refine the trial set of parameters based on the comparison 
and repeat steps (2)-(5) when necessary. We test whether we can 
reach a set of best-fitting parameters that can reproduce 
both the light curve and the apparent spectral evolution as 
observed.

\section{An example: GRB050814}

We apply the method to GRB050814, a typical burst with well-observed
X-ray tail with strong spectral evolution. As seen in
Fig.2, the tail has a steep decay index of $\sim 3.2$, and a strong 
spectral evolution is apparent 
at\footnote{The PC mode spectra become harder at
the end of this tail, which might be due to the contamination of
the harder shallow decay component. For simplification, we focus on 
the WT mode data only.} 
$t<600$ s. These features are common in most of the GRB X-ray tails. 
We first fix $k=1$ in Eq. \ref{eqmain}, which corresponds to the 
simplest cutoff powerlaw model. The initial
trial parameters we choose are ($\Gamma$, $N_{0,p}$, $t_0$,
$E_{p,0}$, $n_{H,host}$) $=(1.2, 0.4, 72.0, 30.0, 0.05)$.
The peak time $t_p$ is fixed to 143.6 s, which corresponds to the
end of the prompt emission.
Some IDL scripts are developed to follow the procedure described 
in \S3 to automatically search for the best-fit parameters  
to match both the observed light curve and the time-dependent 
spectral index. The final best-fitting parameters are shown in Table
1. The corresponding simulated light curve (black curve) and 
spectral indices (green curve) are shown in Fig.2. Figure 2 suggests 
that the sharp decay and the spectral evolution in the tail of GRB 050814
can be indeed explained by the curvature effect with a cutoff power
law spectrum. In Fig.3 we present the comparison between the simualted
and observed spectra in the time steps 1 and 6 (as examples),
which show reasonable consistency.

Our model predicts that the prompt emission spectrum at $t_p \sim 144$
s should be a cut-off power law with the parameters in Table 1. In 
order to confirm this, we subtract the BAT-band spectrum in the time 
interval $(141.5 - 146.5)$ s, and compare with the data with the model
prediction. As shown in Fig.4, the BAT data is roughly consistent with
the model prediction, suggesting the validity of the model.

Some physical parameters can be constrained according to our model.
The time interval from $t_p$ to the beginning of the steep decay phase
$t_{tail,0}$ may be related to the angular spreading time scale
$\tau_{ang}=(t_{tail,0}-t_p)/(1+z)$. Noticing $z \sim 5.3$ for
GRB050814 (Jakobsson et al. 2005), we can estimate the Lorentz factor 
of the fireball as $\Gamma=({R}/2c\tau_{ang})^{1/2}
 \simeq 69 R_{15}^{1/2}$, where $R_{15}=R/(10^{15}~{\rm cm})$ is the
normalized emission radius. Since we
know the spectral peak energy $E_p$ at $t_p$, we can also
estimate the corresponding electrons' Lorentz factor for synchrotron
emission by
$\gamma_{e,p}=\left[E_{p}/({\hbar \Gamma \frac{eB}{mc}})\right]^{1/2}
\sim 2.4\times 10^3 R_{15}^{-1/4}B_3^{-1/2}$. 
From the rest frame duration of the X-ray tail we are analyzing 
$\tau_{tail} = (t_{tail,e} - t_{tail,0})/(1+z) \sim (378-165)/6.3 
= 33.8$ s, one can constrain the 
minimum jet opening angle as $\theta_j > (2c \tau_{tail}/R)^{1/2}
=2.6^{\rm o} \times R_{15}^{-1/2}$. 
These values are generally consistent with those derived from
various other methods.

We find that the abruptness parameter $k$ cannot be very 
different from unity. A Band-function spectrum introduces a 
less significant spectral evolution and it
cannot reproduce the data (cf. Qin 2008b).

\section{Discussions and conclusions}

We have successfully modeled the lightcurve and spectral evolution
of the X-ray tail of GRB050814 using the curvature effect model of
a cutoff power law spectrum with
an exponential cutoff ($k=1$).
It has been
discussed in the literature (e.g. Fan \& Wei 2005; Barniol-Duran
\& Kumar 2008) that the GRB central engine may not die abruptly,
and that the observed X-ray tails may reflect the dying history of
the central engine. If this is indeed the case, the strong spectral 
evolution in the X-ray tails would demand a time-dependent 
particle acceleration mechanism that gives a progressively soft
particle spectrum. Such a behavior has not been predicted by 
particle acceleration theories. Our results suggest that at
least for some tails, the spectral evolution is simply a consequence
of the curvature effect: the observer views emission from the
progressively higher latitudes from the line of sight, so that
the XRT band is sampling the different segments of a curved spectrum.
This is a simpler interpretation.

The phenomenology of the X-ray tails are different from case to case
(ZLZ07). We have applied our model to some other clean X-Ray tails,
such as GRB050724, GRB080523, and find that they can be also interpreted
by this model. Some other tails have superposed 
X-ray flares, making a robust test of the model difficult. A systematic
survey of all the data sample is needed to address what fraction
of the bursts can be interpreted in this way or they demand
other physically distinct models (e.g. Barniol-Duran \& Kumar 2008; 
Dado et al. 2008). This is beyond the scope of this Letter.
  
\acknowledgments
We thank Pawan Kumar for stimulative discussions and comments. This
work is
supported by NASA NNG05GB67G, NNX07AJ64G, NNX08AN24G and NNX8AE57A 
(BBZ\&BZ), by the National Natural Science Foundation of China
under Grant 10463001 (EWL) and 10221001 (XYW), and by the National 
``973'' Program of China under Grant 2009CB824800 (EWL\& XYW). 
BBZ \& BZ also acknowledges the President's Infrastructure Award 
from UNLV.

\end{document}